# TITLE

ML-Net: multi-label classification of biomedical texts with deep neural networks

# AUTHORS:

Jingcheng Du[1,2], B.S.

Qingyu Chen[1], Ph.D.

Yifan Peng[1], Ph.D.

Yang Xiang[2], Ph.D.

Cui Tao[2], Ph.D.

Zhiyong Lu[1*], Ph.D.

[1]National Center for Biotechnology Information (NCBI), National Library of Medicine (NLM), National Institutes of Health (NIH), Bethesda, MD 20894, United States

[2]The University of Texas School of Biomedical Informatics, 7000 Fannin St Suite 600, Houston, TX 77030, United States

[*]corresponding author

# ABSTRACT

*Background*:

In multi-label text classification, each textual document can be assigned with one or more labels. Due to this nature, the multi-label text classification task is often considered to be more challenging compared to the binary or multi-class text classification problems. As an important task with broad applications in biomedicine such as assigning diagnosis codes, a number of different computational methods (e.g. training and combining binary classifiers for each label) have been proposed in recent years. However, many suffered from modest accuracy and efficiency, with only limited success in practical use.

*Methods*:

We propose ML-Net, a novel deep learning framework, for multi-label classification of biomedical texts. As an end-to-end system, ML-Net combines a label prediction network with an automated label count prediction mechanism to output an optimal set of labels by leveraging both predicted confidence score of each label and the contextual information in the target document. We evaluate ML-Net on three independent, publicly-available corpora in two kinds of text genres: biomedical literature and clinical notes. For evaluation, example-based measures such as precision, recall and f-measure are used. ML-Net is compared with several competitive machine learning baseline models.

*Results & Conclusions*:

Our benchmarking results show that ML-Net compares favorably to the state-of-the-art methods in multi-label classification of biomedical texts. ML-NET is also shown to be robust when evaluated on different text genres in biomedicine. Unlike traditional machine learning methods, ML-Net does not require human efforts in feature engineering and is highly efficient and scalable approach to tasks with a large set of labels (no need to build individual classifiers for each separate label). Finally, ML-NET is able to dynamically estimate the label count based on the document context in a more systematic and accurate manner.

The source code of ML-NET is available at: https://github.com/jingcheng-du/ML_Net

# BACKGROUND

Text classification is a common task in natural language processing (NLP) and a building block for many complex NLP tasks. Text classification is the task of classifying an entire text by assigning it one or more predefined labels [1] with broad applications in the biomedical domain, including biomedical literature indexing [2,3], automatic diagnosis codes assignment [4,5], tweets classification for public health topics [6–8], patient safety reports classification [9], etc.

Text classification can be further grouped into two types: 1) multinomial or multi-class text classification that each text is associated with only one of the labels (i.e. labels are mutually exclusive). Binary classification is one of the multinomial classification tasks when only two classes are available; 2) multi-label text classification where each text can be assigned with one or more labels. For example, in Medical Subject Headings (MeSH) indexing, typically a dozen of relevant MeSH terms are assigned to new publications in PubMed [10]. As each textual document can be assigned with indeterminate number of labels, multi-label text classification is often considered harder than multinomial classification [11].

A traditional approach to solving multi-label text classification problem is binary relevance, which decomposes the problem into multiple independent binary classification tasks (one for each label). However, this method assumes the independency of each label [10,12,13]. Label powerset, which creates binary classifiers for each label combination, is able to model potential correlations between labels [14]. However, both of the two approaches could have low throughput when the number of different labels becomes extremely large. There are also some other algorithms for multi-label text classification, including learning to rank [10], classifier chains[15], etc. A review for multi-label learning algorithms can be found [16].

In recent years, deep neural networks have been proposed for multi-label text classification tasks. Most of the efforts [13,17–22] followed a similar framework, which often consists of two modules 1) a neural network that produces scores for each label, including the multi-layer feed-forward neural networks [13,18], the convolution neural networks (CNN) [11,19,20], the recurrent neural networks (RNN) [21], or the ensemble of different types of neural networks [22]. 2) a label predictor that splits the label ranking list into the relevant and irrelevant labels by thresholding methods. However, under this framework, the search for optimal threshold is required and the label decision ignores document context.

Li et al recently incorporated a label decision module into the deep neural networks and achieved the state-of-the-art performance in multi-label image classification tasks [12]. Motivated by their framework, we propose ML-Net, a novel end-to-end deep learning framework, for multi-label biomedical text classification tasks. ML-Net combines the label prediction and label decision in the same network, and is able to determine the output labels based on both label confidence scores and document context. ML-Net aims to minimize pairwise ranking errors of labels, and is able to train and predict the label set in an end-to-end manner, without the need for an extra step to determine the output labels. In order to demonstrate the generalizability and robustness of the deep neural network, we evaluated the framework on three publicly available multi-label biomedical text classification tasks from both biomedical literature domain (two tasks) and clinical domain (one task). We compared the proposed framework with both traditional machine learning baseline models as well as other deep learning models.

# MATERIALS AND METHODS

## DEEP NEURAL NETWORK

The overall architecture of ML-Net can be seen in **Figure 1**. It consists of three major modules: 1) a hierarchical attention network that takes token embedding vectors as the input and outputs the high-dimensional vectors representing the whole textual document; 2) a label prediction network that takes document vectors as the input and outputs the prediction confidence score for each label; 3) a label count prediction network that takes the same document vectors as the input and outputs the estimation of label counts for each document.

### *Text pre-processing*

We first leverage Natural Language Toolkit (NLTK 3.3) to perform the tokenization. We further remove stop words. We then use the pre-trained word embedding [23] to map tokens in the text to high dimensional vectors, which are then fed to the following hierarchical attention networks.

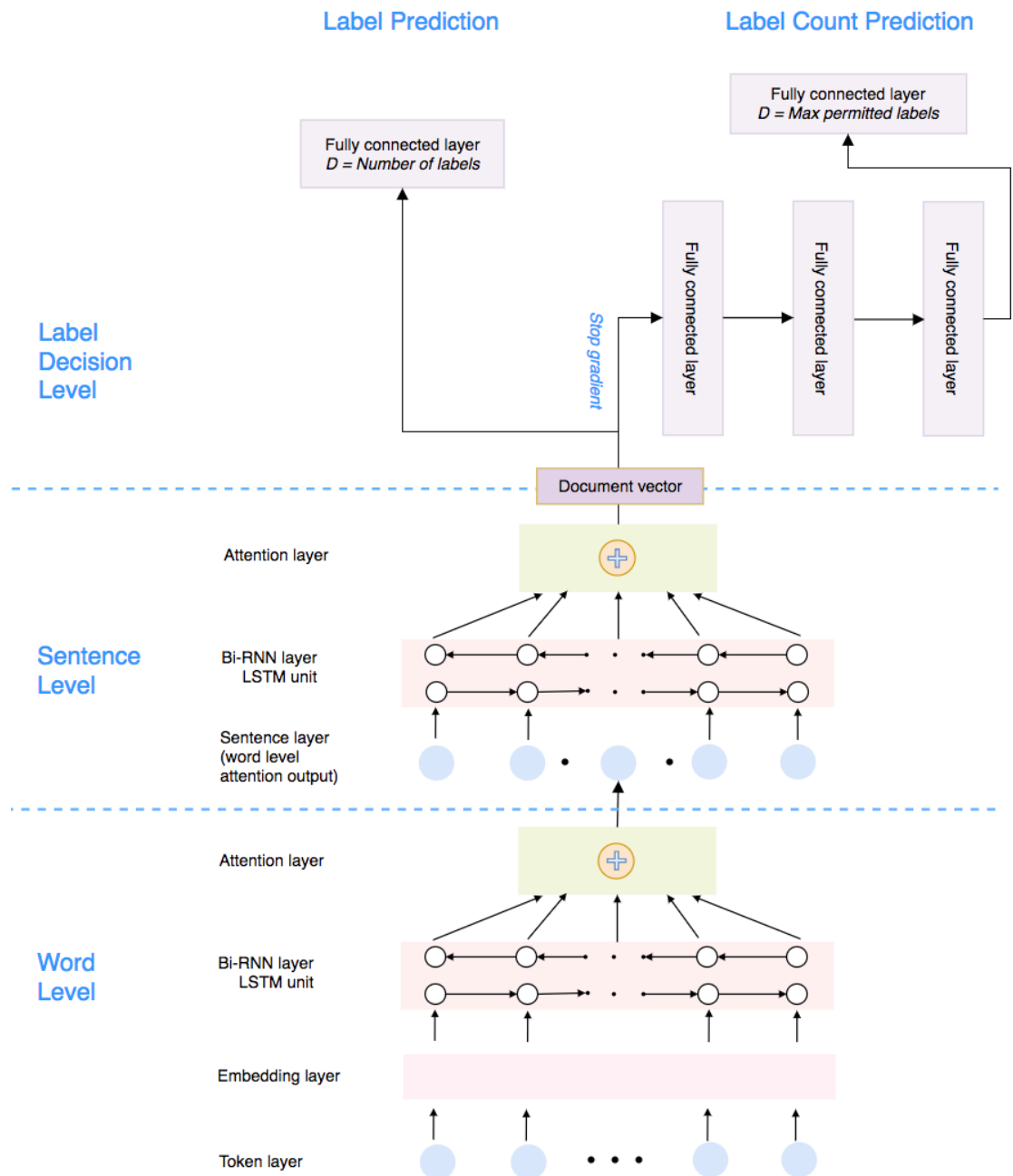

**Figure 1** The framework of ML-Net. The word level takes token embedding as the input and generates representation for each sentence; the sentence takes sentence representation as the input and generates document representation; the label prediction network is a fully connected layer with an output layer that the number of nodes equals to number of unique labels; the label count prediction network consists of three fully connected layers with an output layer that the number of nodes equals to number of maximum permitted labels.

*Hierarchical attention network*

In this study, we apply a hierarchical attention network (HAN) [24] to encode the textual document to high dimensional vectors. This framework has two attention-based encoding levels, see **Figure 1**. The word-level network first takes the word vectors as the input and feed the vectors to a bi-directional RNN, which is able to capture both forward and backward sequential context. We further add the attention mechanism to augment sequence models by capturing the salient portions and context [24,25]. The word level output is further used as the input for the sentence level network. The sentence level network has the same architecture as the word level network.

*Label prediction network*

The label prediction network has a fully connected layer that takes document vector as the input and outputs a predicted confidence score for each label. We apply the rectified linear unit (ReLU) [26] as the activation function for the output. The intuitive objective for multi-label learning is to minimize the number of mis-orderings between the pairs of relevant label and irrelevant labels [13]. Different loss functions have been proposed to model the dependency of individual labels by minimizing the pairwise ranking errors. In this study, we choose log-sum-exp pairwise (LSEP) as our loss function, which has achieved the state-of-the-art performance on large scale multi-label image classification tasks [12]. The equation of LSEP can be seen here:

$$l_{lsep} = \log\left(1 + \sum_{v \notin Y_i} \sum_{u \in Y_i} \exp\left(f_v(x_i) - f_u(x_i)\right)\right)$$

where $f(x)$ is the label prediction function that maps the document vector $x$ into K-dimensional label space representing the confidence scores of each label (K equals to number of unique labels). $f_v(x_i)$ and $f_u(x_i)$ are the $v$ and $u$ -th element of confidence scores for the $i$-th instance in the dataset, respectively. $Y_i$ is the corresponding label set for the $i$-th instance in the dataset.

## *Label count estimation network*

Deciding the proper label set from the predicted label set is a key challenge in multi-label classification. In a common practice, a threshold function is trained to split the ranking of the labels into relevant vs. irrelevant labels [13,18]. However, such thresholding method ignores the document context in decision making. Inspired by a framework from multi-label image classification [12], in this study our label count prediction network takes the document vector as the input and cast the label count estimation as $n$-way classification task, where $n$ is a hyper-parameter for the maximum number of permitted labels that can be returned by the neural network. We design a multilayer perceptron (MLP) network for the label count prediction. This network consists of several fully-connected layers and an output layer with Softmax function for classification.

There are two training steps. We first train the label prediction network. During training, the label prediction as well as the hierarchical attention network are updated through back propagation. Then, we train the label count prediction network. However, different from the training label prediction network, only the MLP part is updated as gradient descent stops at the layer of the document vector. For prediction, we first rank all the individual labels by their corresponding confidence scores generated from the label prediction network, and then the top $K$ (decided by label count prediction network) labels are used as the final output.

## EVALUATION DESIGN

## *Evaluation tasks*

We evaluate our MLT-net on three different text classification tasks with publicly available datasets in two types of text genres: biomedical literature and clinical notes.

Task 1 Hallmarks of cancers classification: The hallmarks of cancers consists of a small number of underlying principles to describe the complexity of cancer[27]. Baker et al introduced a corpus of 1,580 PubMed abstracts manually annotated according to the scientific evidence of 10 currently known hallmarks of cancer [28]. The dataset is available at: https://www.cl.cam.ac.uk/~sb895/HoC.html.

Task 2 Chemical exposure assessments: The vast amount of chemical-specific exposure information available in PubMed is of critical significance. However, the manual collection of such information from PubMed literature can be labor-intensive. Larsoon et al proposed an exposure taxonomy that includes 32 classes, and introduced a corpus of 3,661 abstracts with annotated chemical exposure information [29]. The dataset is available at: https://figshare.com/articles/Corpus_and_Software/4668229 .

Task 3 Diagnosis codes assignment: The automatic assignment of diagnosis codes to medical notes is a useful task, which could benefit computational modeling of patient status. Due to the extremely large label collection, the diagnosis codes assignment task can be considered as an extreme multi-label text classification (XMTC) problem [30]. Perotte et al proposed hierarchy-based classification to automatically assign ICD 9 codes to the discharge summaries from the publicly available Multiparameter Intelligent Monitoring in Intensive Care II (MIMIC II) dataset [4], using hierarchical support vector machine (SVM). We followed the same steps to augment the label set using the hierarchy of the ICD 9 codes in [4]. That is, if an ICD code is in the label set of a document, all of its ancestors are also included in the label set for that document. Their dataset and label augmentation script is publicly available at:

https://physionet.org/works/ICD9CodingofDischargeSummaries .

The overall statistics of the three datasets can be seen in Table 1. Task 1 and 2 have similar characteristics in terms of number of tokens in sentence and number of sentences in document, which is not surprising as they are both collected from PubMed abstracts. In comparison, Task 2 has relative larger number of unique labels and corpus size. The task 3 corpus has a very distinct characteristics than the PubMed abstracts with a significant large number of unique labels (over 7,000) and each document is assigned with many more labels (37 on average, after label augmentation). In addition, unlike PubMed abstracts, each clinical note contains many more short sentences. When we manually review the raw clinical corpus, we find that it is common that a complete sentence in the clinical notes can be split into multiple sentences.

**Table** 1. The description and basic statistics for three tasks. Task 1: hallmarks of cancers classification; task 2: chemical exposure assessments; task 3: diagnosis codes assignment. These are the data after label augmentation. SD: standard deviation

|  | Number of unique labels | Corpus size | Number of tokens in sentence | | | | Number of sentences in document | | | | Number of labels in document | | | |
|---|---|---|---|---|---|---|---|---|---|---|---|---|---|---|
|  |  |  | Mean | Max | Min | SD | Mean | Max | Min | SD | Mean | Max | Min | SD |
| Task 1 | 10 | 1,580 | 15.70 | 83 | 1 | 6.82 | 9.44 | 27 | 2 | 2.87 | 1.56 | 5 | 1 | 0.78 |
| Task 2 | 32 | 3,661 | 16.61 | 120 | 1 | 8.02 | 9.88 | 34 | 1 | 2.81 | 2.05 | 8 | 0 | 1.3 |
| Task 3 | 7,042 | 22,815 | 3.97 | 20 | 1 | 1.90 | 165.71 | 904 | 4 | 95.86 | 36.68 | 127 | 5 | 16.16 |

*Evaluation metric*

The example-based metrics evaluate the multi-label learning system's performance on each test example separately as follows: [16]:

$$Precision = \frac{1}{p}\sum_{i=1}^{p}\frac{|Y_i \cap \hat{Y}_i|}{|\hat{Y}_i|}$$

$$Recall = \frac{1}{p} \sum_{i=1}^{p} \frac{|Y_i \cap \hat{Y}_i|}{|Y_i|}$$

$$F1-score = \frac{2 \cdot Precision \cdot Recall}{Precision + Recall}$$

Where $p$ is the number of instances in the test set. $Y_i$ refers to the true label set for the $i$-th instance in the test set. $\hat{Y}_i$ refers to the predicted label set for the $i$-th instance in the test set.

For task 1 and task 2, we define the true positives as the labels that are identical to the gold-standard labels. For task 3, considering the hierarchical structure of ICD-9 codes, we follow the same definition of true positives in [4], in which true positives are defined as predicted codes that are ancestors of, descendants of, or identical to an assigned code.

### *System implementation*

For task 1 and 2, we used the same implementation as follows. We split the annotated corpus into training, validation and test set with a ratio of 7 : 1 : 2, respectively. The hyper-parameters tuning was performed on the validation set. We used the pre-trained PubMed word2vec (dimension: 200) [31] to map the tokens in the documents to high dimensional vectors. We choose long short-term memory (LSTM) as the RNN unit. We set the number of hidden units in the RNN layer and the dimension of attention output both at 50. Dropout (rate at 0.5) was added on both word level and sentence level output to avoid overfitting. The maximum number of permitted labels was set at 5 and 8 for task 1 and task 2 respectively. The number of neurons in the MLP in the label count prediction network were set at 128, 128, and 64 respectively. We first trained the label prediction network with hierarchical attention network for 50 epochs.

We then applied early stopping while training label count prediction network. We adopt the Adam optimizer [32] and set the learning rate at 0.001.

For task 3, we used the word embedding [33] trained from Multiparameter Intelligent Monitoring in Intensive Care (MIMIC) III corpus [34] using the word2vec algorithm [23]. The dimension was set at 300 empirically. In order to make our model comparable with previous effort, we followed the same data pre-processing steps and used the same datasets for training and testing. We followed almost same hyper-parameters in task 1 and task 2. However, considering the large collection of labels (7,024 unique ICD codes), we set the number of neurons in the MLP of label count prediction network at 7024 (total number of unique labels), 7024, and 128 respectively. The maximum number of permitted labels was set at 70. The major parameters setting for three tasks can be seen in **Table 2**.

Table 2. Major parameters setting in ML-NET for three tasks

| Parameters | Setting | | |
| --- | --- | --- | --- |
| | Task 1 | Task 2 | Task 3 |
| Maximum permitted labels | 5 | 8 | 70 |
| Embedding (dimension) | PubMed word2vec (200) | | MIMIC III word2vec (300) |
| Neurons in the MLP | 128,128,64 | | 7024, 7024, 128 |
| RNN unit (dimension) | LSTM (50) | | |
| Attention layer dimension | 50 | | |
| Dropout rate | 0.5 | | |

| Optimizer (learning rate) | Adam (0.001) |
|---|---|
| Training epochs (label prediction network) | 50 |

Machine learning baseline: For traditional machine learning algorithms, we framed the multi-label classification task as a binary relevance task. We used term frequency–inverse document frequency (TF-IDF) as features and trained a separate binary classifier for each label. We compared multiple machine learning algorithms, including support vector machine (SVM), logistic regression, random forest, extra tress, etc. We report only the results of SVM with learner kernel here as it obtained better performance in all three tasks compared to other algorithms.

Deep learning baseline: we further evaluated other three deep neural networks for these tasks. To assess the effect of the hierarchical attention network, we replaced it in ML-Net with the classic convolutional neural networks proposed by Kim [35] while keeping the label prediction network intact (we call it ML-CNN-Net). In addition, we compared the thresholding methods for the deep neural network models. For ML-NET and ML-CNN-Net, we trained the label prediction network first. Then, we searched the optimal global threshold for the confidence scores generated from label prediction network. The labels, whose confidence scores were higher than the global threshold, were included in the predicted label set. We name these two networks ML-Net-threshold and ML-CNN-Net-threshold respectively. We searched the optimal threshold on the validation set for task 1 and task 2. And for task 3, due to the lack of a validation set, we searched the optimal threshold on the training set.

# RESULTS

The performance of proposed ML-Net and other baseline models is summarized in Table 2. As we can see, ML-Net has the best F-score in both task 1 and task 2. For the hallmarks of cancers task, all the deep learning-based approaches outperformed the binary relevance baseline methods. The ML-Net outperformed the baseline model by over 14%. While for chemical exposure assessments, only the models with proposed label count prediction network (ML-Net and ML-CNN-Net) outperformed the binary relevance baseline methods in F-score. For these two tasks, it is also consistent that the label count prediction network can make better decisions compared to the thresholding methods. In addition, the models with hierarchical attention network achieved a slightly higher F-score than the convolutional neural network in both cases.

For the task of diagnosis codes assignment, the ML-Net with thresholding method achieved better performance than using the label count prediction network. We suspect that this is due to the inclusion of additional codes based on the hierarchical relations in the ICD 9 codes. By doing so, the count of label set might largely depend on the hierarchical structure of ICD 9 codes, instead of the context of the document. As our proposed label count prediction network takes only the document vectors as the input, the label count estimation is not less accurate in this case. Nonetheless, the ML-Net-threshold and ML-CNN-Net-threshold outperformed the binary-relevance baseline, which demonstrated the potentials of deep neural network for the diagnosis codes assignment.

**Table** 3. Comparison of various algorithms for multi-label classification on three tasks. For diagnosis codes assignment, the binary-relevance scores are the best results reported in [4].

|  | Hallmarks of cancers classification | | | Chemical exposure assessments | | | Diagnosis codes assignment | | |
| --- | --- | --- | --- | --- | --- | --- | --- | --- | --- |
|  | Precision | Recall | F-score | Precision | Recall | F-score | Precision | Recall | F-score |
| Binary-relevance (SVM with TFIDF) | 0.742 | 0.688 | 0.714 | **0.778** | 0.677 | 0.724 | **0.577** | 0.300 | 0.395 |
| MT-net | 0.813 | 0.817 | **0.815** | 0.753 | 0.724 | **0.738** | 0.355 | 0.338 | 0.346 |
| MT-net-threshold | 0.752 | **0.837** | 0.793 | 0.700 | 0.735 | 0.717 | 0.492 | 0.360 | 0.416 |
| MT-CNN-net | **0.843** | 0.778 | 0.809 | **0.778** | 0.689 | 0.731 | 0.311 | **0.442** | 0.365 |
| MT-CNN-threshold | 0.764 | 0.817 | 0.790 | 0.662 | **0.774** | 0.713 | 0.501 | 0.373 | **0.428** |

# DISCUSSION & CONCLUSIONS

Compared to the first two tasks, the performance of task 3 is much lower (by all methods) as the task is inherently more challenging. [4] found a slight relationship between diagnosis code prevalence in the training data and performance. The prevalence of diagnosis codes can vary in the corpus. Following label pre-processing steps in [4] and extracting the leaves codes of the augmented ICD codes set, we found that top frequent 100 codes take more than half of total occurrence (115,268 out of 215,805). We also found differences in of diagnosis code co-occurrences in the training and test set. For instance, the counts of the co-occurrence of code '414.01' and 'V45.82' ranks 74[th] in the training set, while ranking 34[th] in the test set; the co-occurrence of code '403.90' and '585.9' ranks 144[th] in the training set, while ranking 12[th] in the test set. The unbalanced codes distribution and the difference of codes co-occurrence in the training and test sets lead to negative impacts on the performance.

When we further examined the difference of prediction and gold standard codes in the test set, we found that the system is more easily to predict diagnosis codes that are close to each other. For example, the count of co-occurrence of codes '412' and '414.01' is 155 in the prediction and 88 in the gold standard; co-occurrence of codes '413.9' and '414.01' is 139 in the prediction and 49 in the gold standard. It is understandable that some closely related codes can be both highly related to the document and thus are included together in the prediction results by the system, while in practice, the nurses or physicians might choose only one from these code pairs.

Compared with binary relevance methods with traditional machine learning algorithms, the proposed deep learning model alleviates human efforts for feature engineering, and avoids building individual classifiers for each label, especially when the label collection is large (e.g. over 1,000 labels). ML-Net advances the state of the art by combining the label prediction network with a label count prediction network, which can not only avoid the manual searching of optimal thresholds for label prediction confidence scores, but also dynamically estimate the label count based on the document context in a more accurate manner. ML-Net achieves the best performance to date in these classification tasks.

Certain limitations remain for this study. Due to the limitations of computation resources, we did not perform a thorough hyper-parameters tuning (i.e. the current parameters setting may not be optimal). In addition, our proposed label count prediction network only takes the document vector as the input. However, the counts of labels might also depend on other information, for example, the hierarchical structure of the labels. Our current network is not able to model such information.

The hierarchical attention network encoding the document can be further improved. Different architecture could be exploited and evaluated. For example, our other studies found that sentence encoder and sentence embedding can provide better representation of the sentences [36,37]. One intuitive change is to replace the word level network with sentence encoder in the hierarchical attention network. In the future, we also plan to apply our proposed network on other larger scale multi-label biomedical text classification tasks, including automatic MeSH indexing, which aims to assign a small set of relevant terms (~12 on average) to a given document from more than 27,000 unique concepts [10].

# DECLARATIONS

*FUNDING*

Research was supported by National Center for Biotechnology (NCBI) Scientific Visitors Program, and Intramural Research Program of the NIH, National Library of Medicine and the Cancer Prevention Research Institute of Texas (CPRIT) Training Grant #RP160015.

*DISCLAIMER*

The content is solely the responsibility of the authors and does not necessarily represent the official views of National Center for Biotechnology, the National Library of Medicine and the Cancer Prevention and Research Institute of Texas.